\documentclass[sigconf]{acmart}
\AtBeginDocument{%
  }
\usepackage{multirow}
\usepackage{booktabs}
\usepackage[most]{tcolorbox} 
\usepackage{hyperref}

\tcbset{colback=gray!10, colframe=black, boxrule=0.5pt, arc=2mm}

\makeatletter
\newcommand{\customlabel}[2]{%
  \protected@write \@auxout {}{\string\newlabel{#1}{{#2}{\thepage}}}%
}
\makeatother

\newtcolorbox[auto counter, number within=section]{errorbox}[1][]{%
  colback=gray!10, colframe=black, fonttitle=\bfseries,
  title=Error examples, #1}

\newcommand{\squishlist}{
  \begin{list}{$\bullet$}{
    \setlength{\itemsep}{0pt}
    \setlength{\parsep}{3pt}
    \setlength{\topsep}{3pt}
    \setlength{\partopsep}{0pt}
    \setlength{\leftmargin}{1.5em}
    \setlength{\labelwidth}{1em}
    \setlength{\labelsep}{0.5em}
  }
}

\newcommand{\squishlisttwo}{
  \begin{list}{$\bullet$}{
    \setlength{\itemsep}{0pt}
    \setlength{\parsep}{0pt}
    \setlength{\topsep}{0pt}
    \setlength{\partopsep}{0pt}
    \setlength{\leftmargin}{2em}
    \setlength{\labelwidth}{1.5em}
    \setlength{\labelsep}{0.5em}
  }
}

\newcommand{\squishend}{
  \end{list}
}

\usepackage{float}

\usepackage{xcolor} 

\setcopyright{acmlicensed}
\copyrightyear{2018}
\acmYear{2018}
\acmDOI{XXXXXXX.XXXXXXX}
\acmConference[Conference acronym 'XX]{Make sure to enter the correct
  conference title from your rights confirmation email}{June 03--05,
  2018}{Woodstock, NY}
\acmISBN{978-1-4503-XXXX-X/2018/06}

\begin{document}

\title{Smart Trial: Evaluating the Use of Large Language Models for Recruiting Clinical Trial Participants via Social Media}
\author{Xiaofan Zhou}

\authornotemark[1]
\email{xzhou77@uic.edu}
\affiliation{%
  \institution{University of Illinois at Chicago}
  \city{Chicago}
  \state{Illinois}
  \country{USA}
}

\author{Zisu Wang}
\affiliation{%
  \institution{Colorado State University}
  \city{Fort Collins}
  \state{Colorado}
  \country{USA}}
\email{zisu.wang@colostate.edu}

\author{Janice Krieger}
\affiliation{%
  \institution{Mayo Clinic}
  \city{Jacksonville}
  \state{Florida}
  \country{USA}
}
\author{Mohan Zalake}
\email{zalake@uic.edu}
\affiliation{%
  \institution{University of Illinois at Chicago}
  \city{Chicago}
  \state{Illinois}
  \country{USA}
}
\author{Lu Cheng}
\email{lucheng@uic.edu}
\affiliation{%
  \institution{University of Illinois at Chicago}
  \city{Chicago}
  \state{Illinois}
  \country{USA}
}
\begin{abstract}
Clinical trials (CT) are essential for advancing medical research and treatment, yet efficiently recruiting eligible participants--each of whom must meet complex eligibility criteria--remains a significant challenge. Traditional recruitment approaches, such as advertisements or electronic health record screening within hospitals, are often time-consuming and geographically constrained. This work addresses the recruitment challenge by leveraging the vast amount of health-related information individuals share on social media platforms. With the emergence of powerful large language models (LLMs) capable of sophisticated text understanding, we pose the central research question: \textit{Can LLM-driven tools facilitate CT recruitment by identifying potential participants through their engagement on social media?} To investigate this question, we introduce \textbf{TRIALQA}, a novel dataset comprising two social media collections from the subreddits on colon cancer and prostate cancer. Using eligibility criteria from public real-world CTs, experienced annotators are hired to annotate TRIALQA to indicate (1) whether a social media user meets a given eligibility criterion and (2) the user’s stated reasons for interest in participating in CT. We benchmark seven widely used LLMs on these two prediction tasks, employing six distinct training and inference strategies. Our extensive experiments reveal that, while LLMs show considerable promise, they still face challenges in performing the complex, multi-hop reasoning needed to accurately assess eligibility criteria.
\end{abstract}



\received{20 February 2007}
\received[revised]{12 March 2009}
\received[accepted]{5 June 2009}

\maketitle

\section{Introduction}

Clinical trials (CT) are essential for advancing medical knowledge, ensuring that new treatments, drugs, and interventions are safe and effective for human use \cite{zhang2025importance}. An important topic in medical research is the recruitment of individuals who meet specific CT criteria and show interest in participating in the CT. However, identifying suitable candidates remains a significant challenge, with an estimated 80\% of trials failing to meet their enrollment targets on time, and 55\% of terminated trials are due to poor accrual \cite{clinicaltrialsarena2012patientrecruitment}. These shortfalls result in underpowered studies that undermine the validity and generalizability of biomedical research. Traditional recruitment methods, such as direct outreach, mailings, and email campaigns, often fall short in efficiently reaching large populations \cite{heller2014strategies, friedman2013people}. This work addresses a major problem that \textit{existing approaches do not make use of the vast amount of health-related information people actively share on social media platforms to identify individuals for CT participation.} Although forums like Reddit \footnote{https://www.reddit.com} are widely used to advertise CT, these efforts typically rely on generic targeting and overlook rich engagement signals, such as posts, that could reveal users’ health concerns, interests, and potential eligibility. 

The emerging powerful LLMs such as LLaMA \cite{Touvron2023LLaMA}, DeepSeek \cite{deepseekai2025deepseekr1incentivizingreasoningcapability}, Claude \cite{anthropic2024claude3, anthropic2025claude4}, and GPT-4 \cite{openai2023gpt4} have demonstrated strong performance across a range of tasks, including question answering \cite{sun2018open, kaiser2024robust}, summarization \cite{paulus2017deep, moradan2025untapping}, and mathematical problem solving \cite{wei2022chain, yang2024gap}. Recent research has introduced a variety of techniques to better harness the capabilities of LLMs--such as chain-of-thought reasoning (CoT)\cite{wei2022chain, chiang-etal-2025-tract, yu2025explainable}, retrieval-augmented generation (RAG)\cite{lewis2020retrieval, wu2025lighter}, self-consistency\cite{wang2022self, huang-etal-2024-enhancing}, and few-shot prompting\cite{brown2020language,wei2021finetuned, alsuhaibani2024idofew}--which have been shown to be effective across diverse domains. Given that social media posts can be lengthy, including users' history and interactions, a promising solution for the CT recruitment problem is to leverage LLMs to identify potential participants from social media platforms such as Reddit, where users openly discuss health issues in specialized subreddits, providing rich, self-reported data. Our overarching research question in this work is: 

\textit{\textbf{Can LLM-driven tools be used for CT recruitment by identifying potential participants through online users' engagement on social media?}}

To answer this question, we first carefully curate \textit{Trial-QA,} including two social media datasets collected from Reddit that capture individuals’ alignment with CT criteria for colon cancer study and prostate cancer study. The colon cancer dataset consists of 502 Reddit users collected from the colon cancer subreddit, 8 CT criteria for colon cancer identified from public CT recruitment, and 6 indicators of research interest reasons in colon cancer identified by the CT expert. The prostate cancer dataset consists of 799 Reddit users collected from the prostate cancer subreddit, and 8 CT criteria for prostate cancer identified from the corresponding public CT recruitment. Both datasets are annotated by experienced annotators from a medical or computer science background. \textit{Trial-QA} is designed to evaluate the ability of LLMs to identify potential CT participants from social media platforms. We then assess the performance of 7 widely used LLMs, ranging from small to large-scale models, using 5 different prompting and reasoning methods. This work lays the foundation for future research in automating and improving the recruitment process for CTs.

Our contributions can be described as:
\squishlist
    \item We pioneer the study of \textit{smart trial recruitment}, which leverages the vast amount of health-related information that people actively share on social media platforms and emerging powerful LLMs to identify individuals for CT participation. 
    \item We collect Reddit posts related to colon and prostate cancer and construct the \textsc{TrialQA} dataset, where each post is annotated with multiple eligibility criteria and potential reasons for research interest. To the best of our knowledge, this is the first dataset specifically designed to evaluate the use of LLMs for CT participant recruitment.
    \item We evaluate a range of LLMs of varying sizes and provide a detailed analysis of their performance across aligning user posts with different criteria and interest types. Our main finding is that current LLMs still struggle with this task, indicating significant room for improvement in CT eligibility reasoning.
\squishend

\begin{figure}
    \centering
    \includegraphics[width=.9\linewidth]{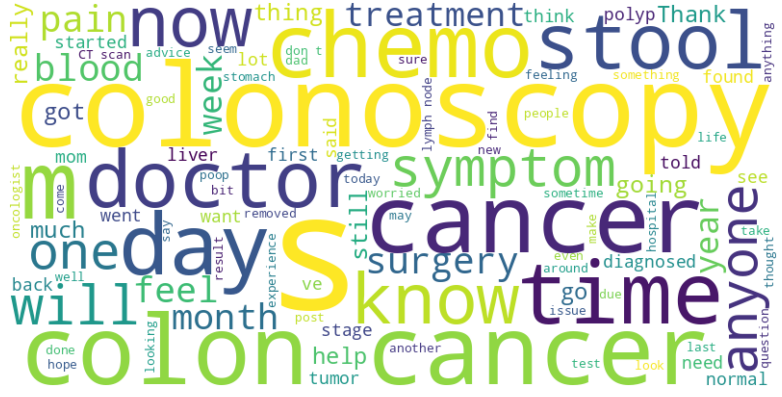}
    \caption{Word cloud of colon cancer posts}
    \label{colon_word}
\end{figure}
\begin{figure}
    \centering
    \includegraphics[width=.9\linewidth]{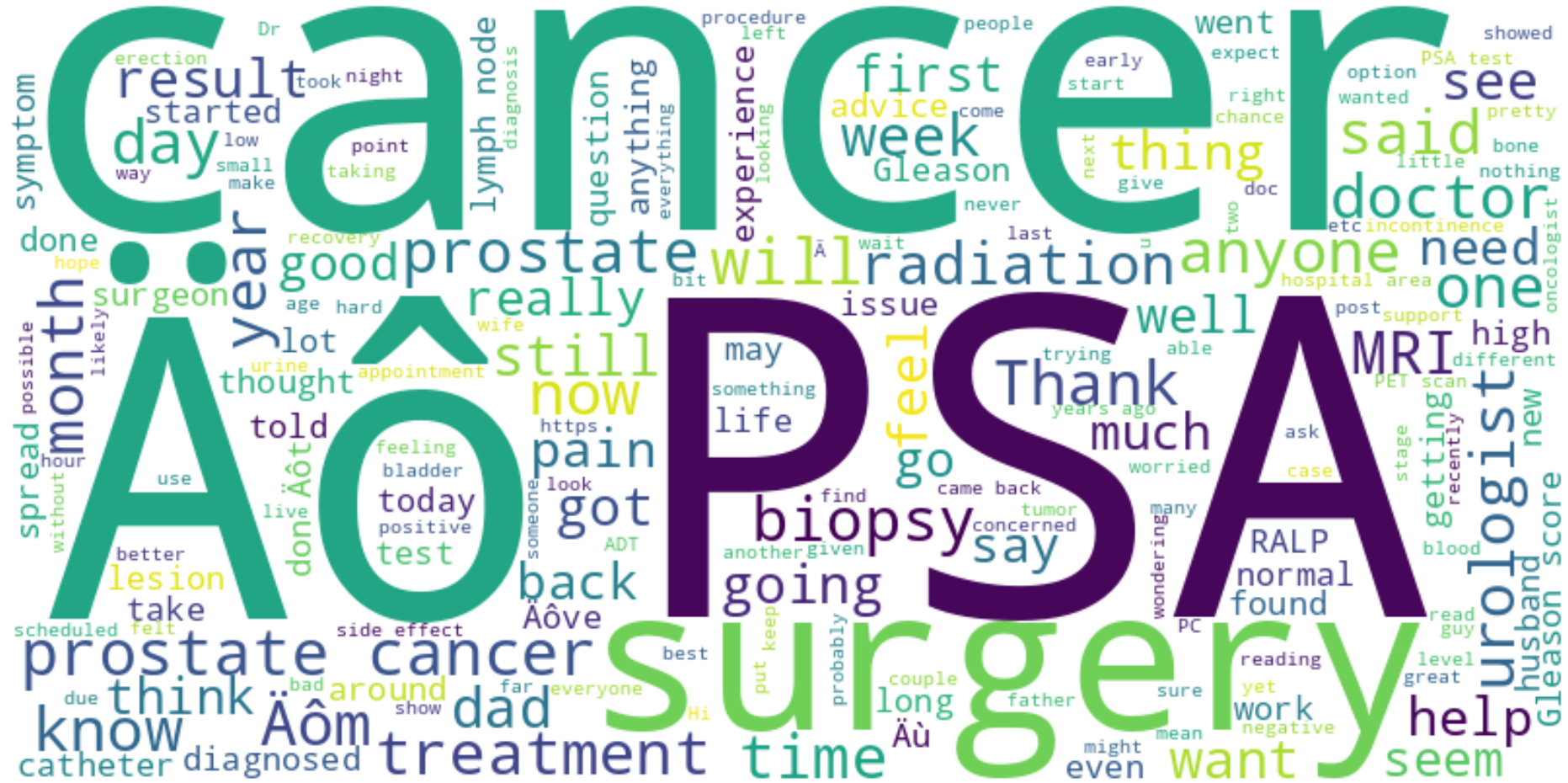}
    \caption{Word cloud of prostate cancer posts}
    \label{pro_word}
\end{figure}
\section{TrialQA -- A social media dataset for online CT recruitment} 

To evaluate LLMs' ability of predicting social media user's clinical eligibility criteria and potential reasons for participating in the CT, we need to annotate each post with the corresponding labels for every criterion and interest reason. 
Below, we present the creation of our datasets, TRIALQA, designed specifically for participant recruitment based on the eligibility criteria of the Metabiomics Colon Cancer Clinical Research Study\footnote{\url{https://clinicaltrials.gov/study/NCT02151123}} and the Prostate Cancer Screening for People at Genetic Risk for Aggressive Disease\footnote{\url{https://clinicaltrials.gov/study/NCT04472338}}. We first describe the dataset curation process, followed by a detailed analysis. To the best of our knowledge, these are the \textbf{first} annotated datasets focused on classifying potential CT participants for recruitment purposes on social media.

\subsection{Social Media Post Collection}

\begin{table}[ht]
\centering
\caption{Basic statistics for the \texttt{r/coloncancer} and \texttt{r/prostatecancer} datasets.}
\label{tab:subreddit_stats}
\resizebox{1.0\linewidth}{!}{
\begin{tabular}{lcc}
\hline
 & \texttt{r/coloncancer} & \texttt{r/prostatecancer} \\
\hline
\textbf{Sample size (users)} & 502 & 799 \\
\textbf{Sample size (posts)} & 806 & 1,555 \\
\textbf{Average posts per user} & 1.61 (range: 1--15) & 1.94 (range: 1--60) \\
\textbf{Average words per user} & 241.70 $\pm$ 266.70 & 290.80 $\pm$ 445.27 \\
\textbf{Min words number} & 8 & 12 \\
\textbf{Max words number} & 3,232 & 5,695 \\
\textbf{1st quartile (Q1) of words number} & 88 & 83 \\
\textbf{Median of words number} & 164.5 & 158 \\
\textbf{3rd quartile (Q3) of words number} & 310 & 308 \\
\hline

\end{tabular}
}
\end{table}

We collect all posts publicly available on \texttt{r/coloncancer} and \texttt{r/prostatecancer} subreddits. 
For each post, we retrieve its title, content, URL, creation time, update time, and creator's ID. 
Due to the large number of posts, we randomly select a sample of 502 users and the corresponding 806 posts for \texttt{r/coloncancer}, while we randomly select a sample of 799 users and the corresponding 1,555 posts for \texttt{r/prostatecancer}. For each selected user, we concatenate all of their posts in chronological order to reconstruct the user's post history, which serves as the input context for annotation and analysis. We show the basic statistics of these two datasets in Table \ref{tab:subreddit_stats}. In both datasets, a high standard deviation of words per user suggests marked variability in user verbosity; moreover, the data analyses results indicate that word-frequency distributions are highly skewed in both subreddits.
Additionally, the majority of users posted only once, indicating limited engagement per user. However, a subset of users exhibited markedly higher activity levels, contributing multiple posts within a single context entry. 


To evaluate the viability of using user-generated content for CT recruitment, we analyzed these Reddit posts and visualized keyword frequency using a word cloud (Figure~\ref{colon_word} and Figure~\ref{pro_word}). The visualizations reveal frequent use of medically relevant terms. For \texttt{r/coloncancer} data, keywords such as ``colonoscopy'', ``chemo'' and ``colon cancer'' are prominent, while posts from \texttt{r/prostatecancer} data show frequent mentions of ``PSA'', ``surgery'', ``radiation'' and ``prostate''. These terms indicate that users are actively discussing diagnostic procedures, treatment history, and disease management—topics that closely mirror CT eligibility criteria.

The presence of such content suggests that many users voluntarily share information that is highly relevant for CT screening. For example, posts that include words like ``doctor'', ``time'' or ``surgery'' provide direct or inferable information about doctor diagnosis, treatment initiation, or surgery for illness. In addition, references to age-related terms such as ``years'', ``young'' or ``old'' may help estimate age eligibility, even when not explicitly stated. This kind of informal but context-rich language offers a practical signal for identifying individuals who may meet inclusion or exclusion criteria.


Moreover, users often express uncertainty or seek advice using language such as ``help'', ``anyone'' or ``advice''. These expressions reflect a potential openness to CT research, especially when users are navigating treatment decisions or exploring new options. This help-seeking behavior is valuable from a recruitment perspective, as it may indicate greater willingness to consider trial participation when informed and supported.

Together, these observations demonstrate that Reddit posts offer more than just anecdotal discussion—they contain medically relevant, self-disclosed information embedded in natural language. This makes them a promising and scalable resource for identifying potential participants and prioritizing outreach for CT recruitment.

\subsection{Criteria and Interest Reason Design}
To identify eligible individuals for a colon cancer CT on Reddit, we adopt the eligibility criteria from a real-world CT study that was conducted in Colorado, the United States, and launched in 2018. The eligibility criteria include a set of inclusion and exclusion criteria used to recruit eligible CT participants. On the recommendation of a colon cancer clinical expert at the Mayo Clinic\footnote{https://www.mayoclinic.org/}, we further designed six interest reason indicators, indicating the specific reasons that a social media user may be interested in participating in the colon cancer CT:

\squishlist
\item r1: Searching for research studies related to colon cancer;
\item r2: Searching for clinical studies on colon cancer;
\item r3: Searching for investigations concerning colon cancer;
\item r4: Searching for new medications for colon cancer;
\item r5: Searching for new therapies for colon cancer;
\item r6: Searching for information about colon cancer CT.
\squishend

If a user's context contains any of these indicators, they are considered potentially interested int participating in the CT. 


Four inclusion criteria that a user must satisfy are:
\squishlist
\item c1: Recently diagnosed with colorectal cancer (CRC) and scheduled for colectomy;
\item c2: Age between 18 and 95 years;
\item c3: Able to comprehend, sign, and date the written informed consent form (ICF);
\item c4: Able to provide informed consent in English.
\squishend
Additionally, four exclusion criteria are:
\squishlist
\item c5: History of Inflammatory Bowel Disease (IBD);
\item c6: Use of antibiotics within two weeks prior to sample collection;
\item c7: Colonoscopy, colon preparation, or bowel contrast agent within seven days prior to sample collection;
\item c8: History of any radiation therapy.
\squishend
For the prostate cancer dataset, we adopt the eligibility criteria from a prostate cancer screening study for individuals at genetic risk of aggressive disease. c8 (see below) in the prostate cancer criteria includes a requirement to display the interest reasons for participating in the CT; thus, we do not annotate interest reason indicators. Two inclusion criteria are

\squishlist
\item c1: People with prostates $\geq$40 years of age;
\item c2: Documented germline pathogenic variant in known or suspected genes associated with prostate cancer risk;
\squishend
And six exclusion criteria are:
\squishlist
\item c3: Prior diagnosis of prostate cancer;
\item c4: Medical contraindication to any of the study procedures (e.g., prostate biopsy);
\item c5: For all cancer types except non-melanoma skin cancer, any cancer treatment with curative intent within the past 12 months (e.g., surgery, radiation, chemotherapy, immunotherapy);
\item c6: Prior or concurrent participation in an interventional CT aimed at preventing cancer for people with germline variants associated with increased prostate cancer risk;
\item c7: Unable to provide written informed consent;
\item c8: Unable or unwilling to complete clinical care and study procedures as indicated by the study protocol.;
\squishend

Overall, we select individuals who meet all inclusion criteria, do not meet any exclusion criteria, and exhibit at least one indicator of potential interest in participating in the CT. 
\subsection{Annotation Protocol}
We hire three experienced annotators, two from computer science (CS) background and one from medicine background. They were provided with all user posts, along with the eligibility criteria and interest reason indicators, and were tasked with determining whether each user met the corresponding requirements based on the content of their posts. All annotators are fluent in English and trained in CT annotation. The two annotators from CS conduct a first round of annotation, and whenever a disagreement arises, the annotator from the medicine background then help reconcile the annotation results.
Each eligibility criterion was annotated using a three-class scheme: ``True'' if the post clearly indicated that the user met the criterion, ``False'' if the post indicated that the user did not meet the criterion, and ``Unknown'' if the relevant information was not mentioned. For each interest reason indicator, a binary label was used: ``True'' if the post showed that the user was interested in participating in a CT, and ``False'' otherwise.

\subsection{Annotation Results Analysis}
\begin{figure}
    \centering
    \includegraphics[width=1\linewidth]{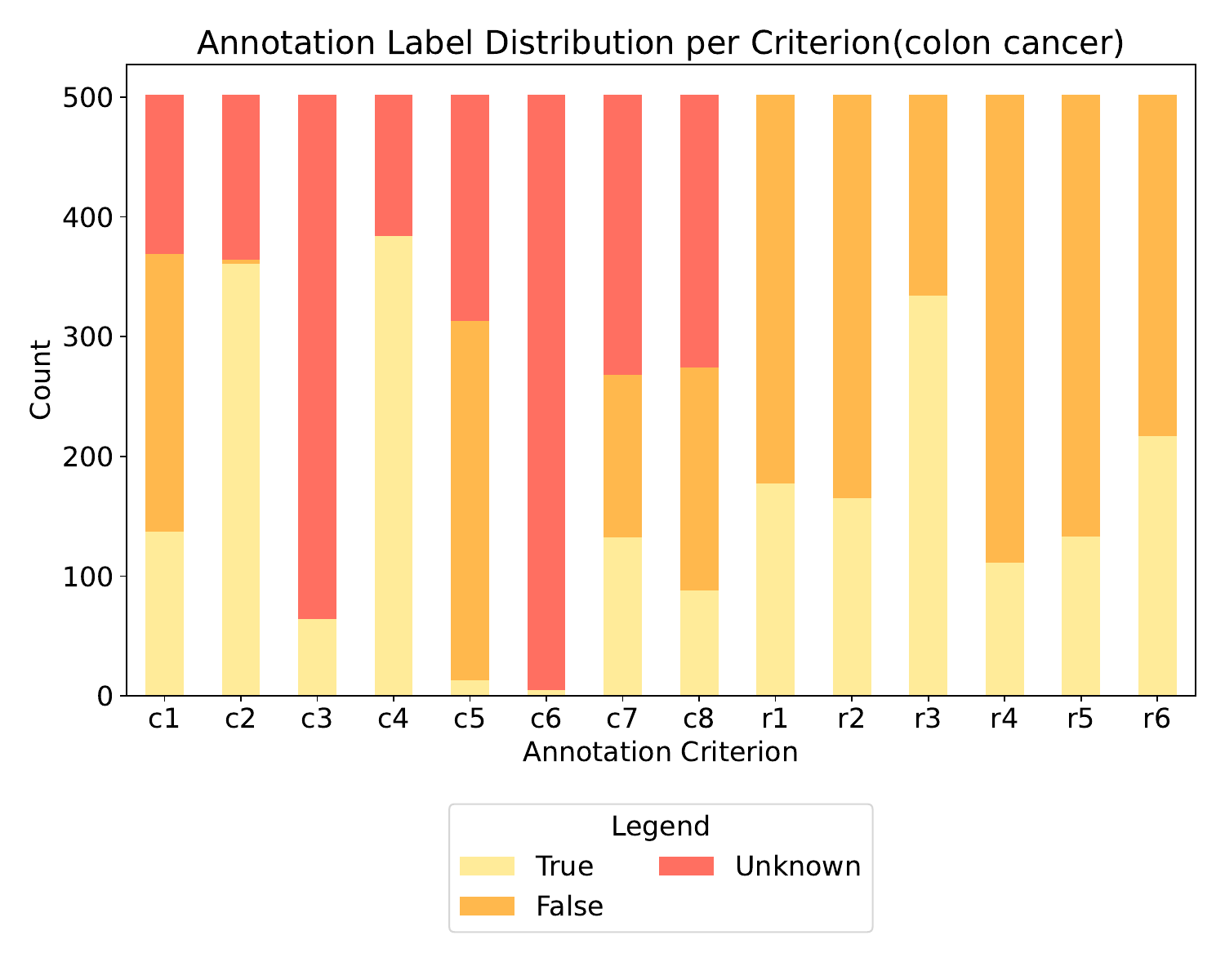}
    \caption{Distribution of colon cancer CT criteria annotation.}
    \label{dis_colon}
\end{figure}
\subsubsection{Colon Cancer Data}
The annotation results in Figure~\ref{dis_colon} for colon cancer CT criteria (c1–c8) reveal varied levels of eligibility among users. Criteria such as c2 (age between 18 and 95) and c4 (ability to provide informed consent in English) have a substantial number of ``True'' labels. These findings suggest that such characteristics are relatively common in the user population, making them potentially strong filters for identifying eligible participants in CT. In contrast, criteria like c1 (diagnosed with colorectal cancer and scheduled for colectomy) and c5 (history of IBD) are more frequently labeled ``False'' indicating that these conditions are less commonly reflected in user posts. 

A considerable number of ``Unknown'' labels appear in criteria such as c3 (Ability to sign informed consent) and c6 (Recent antibiotic use). These criteria often depend on information that is not explicitly provided in social media posts, highlighting the difficulty of evaluating certain eligibility aspects from user-generated content and underscoring the need for additional context or external information to improve CT eligibility assessment from online discussions.

For the colon cancer research interest reasons (r1–r6), there is a relatively balanced distribution between the ``True'' and ``False'' labels. This suggests that many individuals express at least one reason for being interested in participating in a CT. These users may be more receptive to recruitment messaging or engagement with research opportunities. The distribution of these interest signals demonstrates the potential value of leveraging online forums to identify and connect with motivated participants.

\begin{figure}
    \centering
    \includegraphics[width=1\linewidth]{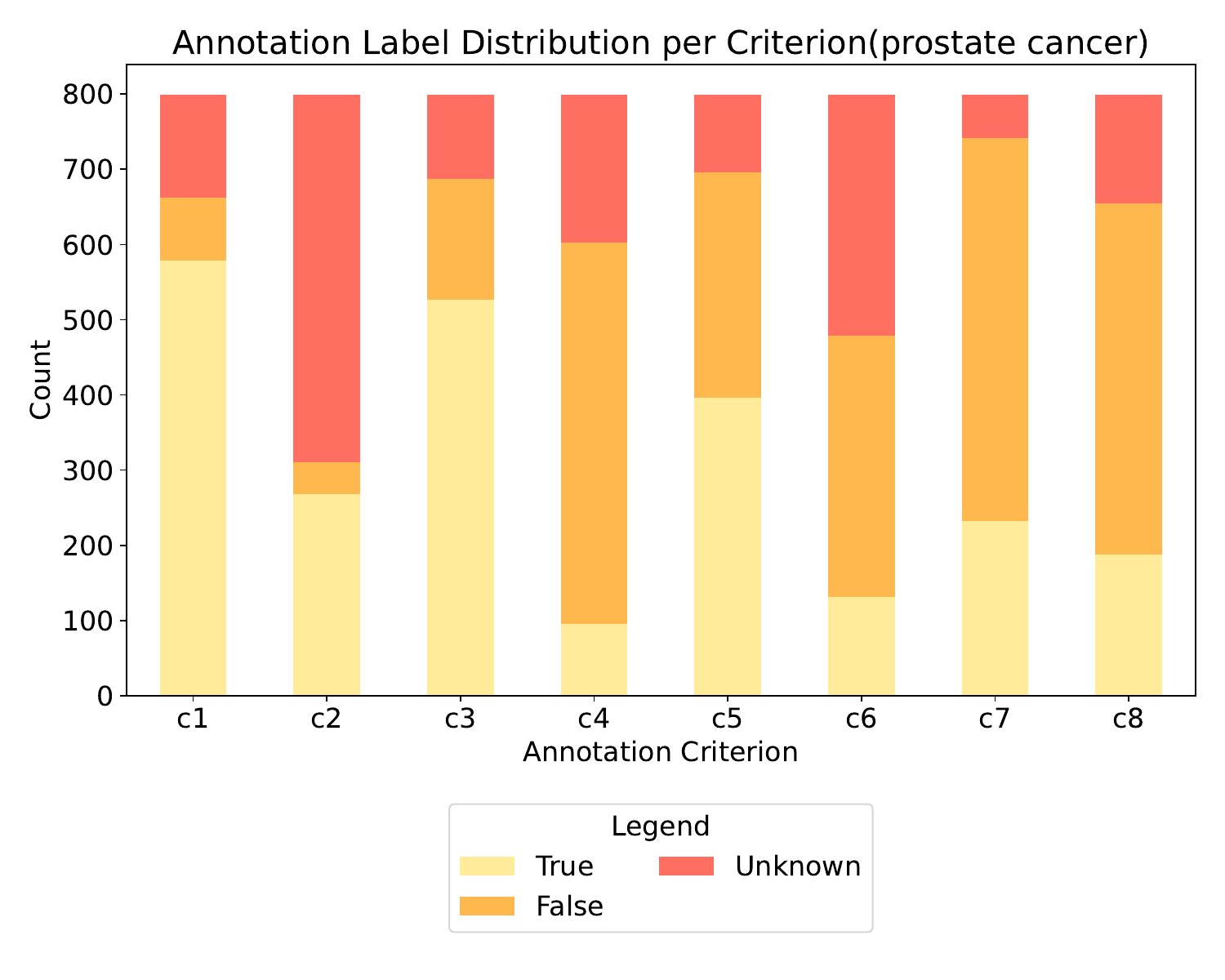}
    \caption{Distribution of prostate cancer CT criteria annotation.}
    \label{dis_pro}
\end{figure}
\subsubsection{Prostate Cancer Data}
The annotation results in Figure~\ref{dis_pro} for prostate CT criteria (c1–c8) reveal clear patterns across both inclusion and exclusion conditions. We find that the distribution in this dataset is relatively balanced, with some imbalance in c1 (age $\geq$40), c2 (Pathogenic germline variant linked to prostate cancer risk), and c7 (Unable to provide written informed consent). This may be explained by the fact that c1 reflects the subreddit’s older demographic, c2 is rarely mentioned because individuals without prostate cancer–related genetic risk often do not disclose such information, and c7 is straightforward to determine given the ease of assessing the ability to provide written informed consent. Overall, this dataset has relatively balanced labels.

 
\section{Experiments}
To answer our overarching research question ``can LLM-driven tools be used for CT recruitment by identifying potential participants through online users' engagement on social media?'' we conduct a series of experiments on the TRIALQA dataset using various LLM architectures with different training and inference strategies. Specifically, we aim to answer the following five research questions: \textbf{RQ1}: How do different LLM training and inference strategies perform on our smart trial recruitment task? \textbf{RQ2}: How do different LLM strategies perform on different eligibility criteria and reasons for interest? \textbf{RQ3}: How do different LLMs perform on different classes, i.e., \{True, False, Unknown\}? \textbf{RQ4}: Can the popular CoT strategy used to induce LLM reasoning improve LLMs' performance? \textbf{RQ5}: What are the common types of errors made by these LLMs?

\subsection{Experimental Setup}
LLMs vary in their capabilities across different tasks, and model size can significantly influence performance. To explore these effects, we evaluate seven open-source LLMs: three large models ($\geq$ 70B parameters) and four smaller models (7–8B parameters). The large models are Llama3-70B (Meta-Llama-3-70B-Instruct) \cite{llama3modelcard}, Qwen2.5-72B (Qwen2.5-72B-Instruct) \cite{qwen2,qwen2.5}, and DeepSeek-L70B (DeepSeek-R1-Distill-Llama-70B) \cite{deepseekai2025deepseekr1incentivizingreasoningcapability}. The smaller models are Llama3.1-8B (Llama-3.1-8B-Instruct) \cite{llama3_2024}, Qwen2.5-7B (Qwen2.5-7B-Instruct) \cite{qwen2,qwen2.5}, DeepSeek-Q7B (DeepSeek-R1-Distill-Qwen-7B) \cite{deepseekai2025deepseekr1incentivizingreasoningcapability}, and Mistral-7B (Mistral-7B-Instruct-v0.3) \cite{mistral7b_2023}. For comparison with a non-LLM baseline, we include RoBERTa (Roberta-large-mnli) \cite{liu2019roberta}, a transformer-based model fine-tuned for natural language inference (NLI) that classifies premise–hypothesis pairs as True, False, or Unknown, which aligns well with our task. 

\noindent\textbf{Metrics.}
To evaluate the performance of LLMs, we use common classification metrics, including accuracy (acc), macro F1 score (MF1), and weighted F1 score (WF1). Accuracy reflects the overall proportion of correct predictions. Macro-F1 treats all classes equally, making it suitable for assessing performance on minority classes in imbalanced datasets, as in our case. Weighted-F1 accounts for class frequency, providing a distribution-aware performance measure. Using these three metrics together offers a comprehensive view of both overall and per-class model performance. We use regular expressions to extract the label from the prediction.

\subsection{Approach}
We consider six distinct and widely used LLM training and inference strategies. First, we include \textbf{direct LLM inference}, where the model generates predictions without additional contextual guidance. We use the following prompt for direct LLM inference:
\begin{tcolorbox}[colback=gray!10, colframe=gray!80]
What is the answer to the question based solely on the information in the sentence? Focus only on the sentence content. Respond with true, false, or unknown. Use unknown if the sentence does not provide enough information to answer.
Sentence: \{User posts\}.
Questions: \{whether the user meets a criterion or has a reason for interest?\}
Answer:
\end{tcolorbox}
Second, we examine \textbf{in-context learning (ICL)} \cite{brown2020language}, in which few-shot examples are provided in the prompts during inference to guide the model's reasoning. We use the following prompt for ICL:
\begin{tcolorbox}[colback=gray!10, colframe=gray!80]
What is the answer to the question based solely on the information in the sentence? Focus only on the sentence content. Respond with true, false, or unknown. Use unknown if the sentence does not provide enough information to answer.
\{Examples\}
Sentence: \{User posts\}.
Questions: {whether the user meets a criterion or has a reason for interest?}
Answer:
\end{tcolorbox}
Third, we assess the \textbf{self-consistency }method (consist) \cite{wang2022self}, which generates multiple outputs for a given input, and the final answer is the most consistent output. 

In addition, we explore \textbf{chain-of-thought} (CoT) reasoning \cite{wei2022chain}, where the model is prompted to articulate intermediate reasoning steps before reaching a final decision, with the goal of improving accuracy on complex criteria. We use a two-step strategy to induce reasoning where the second step is used for get a answer following instruction from reasoning. In the first step, we prompt the LLM to generate its reasoning:
\begin{tcolorbox}[colback=gray!10, colframe=gray!80]
What is the answer to the question based solely on the information in the sentence? Explain your answer step by step.
Sentence: \{User posts\}.
Questions: \{whether the user meets a criterion or has a reason for interest?\}
Answer:
\end{tcolorbox}
After obtaining the reasoning, we prompt the LLM to produce the final answer based on it:
\begin{tcolorbox}[colback=gray!10, colframe=gray!80]
Based on reasoning provided by a different LLM for the user's post, answer the question with true, false, or unknown.
Example \{Examples\}
Now answer the following and return one of true, false, or unknown directly at the beginning of your answer:
Sentence: \{User posts\}.
Reasoning: \{Reasoning\}.
Questions: \{whether the user meets a criterion or has a reason for interest?\}
Answer:
\end{tcolorbox}
In addition to various inference strategies, we also evaluate LLMs' fine-tuning under two different settings: User-level fine-tuning (User-level FT) and Entry-level fine-tuning (Entry-level FT.
In the User-level FT setting, each user is treated as a single sample
We feed LLMs all historical posts from a single user, and the output is a set of eligibility criteria and interest reason labels associated with that user. 
In the Entry-level FT setting, each post is treated as a single sample: $(\text{post}, \text{criterion label})$ or $(\text{post}, \text{interest reason label})$. The entry-level setting is used in all inference strategies.
Finetuning in both settings is implemented using LoRA \cite{hu2022lora}. This comprehensive comparison offers insights into the relative strengths and limitations of different LLM-based techniques for CT participant recruitment. The finetuning for the eligibility criteria and interest reason predictions is conducted separately as they have different number of class. We split training data and test data by 4:1. For the non-LLM Roberta, we use an NLI model, treating the user post as the premise and the criterion or research interest reason as the hypothesis. The model outputs True, False, or Unknown based on the alignment between the user post and the criterion or research interest. The results for Roberta are reported under the ``Direct Prompt'' column for comparison.
\begin{table*}[]
\centering
\caption{Main results for different LLM inference strategies on TRIALQA.}
\begin{tabular}{lc ccc ccc ccc}
\toprule
\textbf{Dataset} &\textbf{Model} & \multicolumn{3}{c}{\textbf{Direct}} & \multicolumn{3}{c}{\textbf{ICL}} & \multicolumn{3}{c}{\textbf{Consist}}\\
\cmidrule(r){3-5} \cmidrule(r){6-8} \cmidrule(r){9-11}
 & & acc & MF1 & WF1 & acc & MF1 & WF1 & acc & MF1 & WF1 \\
\midrule

\multirow{9}{*}{Colon cancer} 
& Roberta & 0.55 & 0.49 & 0.49 & - & - & - & - & - & - \\
\cmidrule{2-11}
& Llama3-70B & 0.52 & 0.48 & 0.48 & 0.56 & 0.53 & 0.54 & 0.55 & 0.51 & 0.52 \\
& Qwen2.5-72B  & 0.57 & 0.50 & 0.50 & 0.57 & 0.50 & 0.50 & 0.57 & 0.50 & 0.50 \\
& DeepSeek-L70B & 0.47 & 0.44 & 0.46 & 0.55 & 0.53 & 0.53 &  0.54 & 0.52 & 0.53 \\
\cmidrule{2-11}
& Llama3.1-8B & 0.53 & 0.51 & 0.51 & 0.51 & 0.48 & 0.49 & 0.50 & 0.47 & 0.48 \\
& Qwen2.5-7B & 0.56 & 0.49 & 0.49 & 0.56 & 0.47 & 0.48 & 0.54 & 0.45 & 0.45 \\
& DeepSeek-Q7B & 0.53 & 0.52 & 0.51 & 0.56 & 0.48 & 0.48 & 0.55 & 0.46 & 0.46\\
& Mistral-7B & 0.58 & 0.52 & 0.52 & 0.58 & 0.57 & 0.57 & 0.58 & 0.57 & 0.58\\

\cmidrule{1-11}
\multirow{9}{*}{Prostate cancer} 
& Roberta & 0.36 & 0.33 & 0.32 & - & - & - & - & - & - \\
\cmidrule{2-11}
& Llama3-70B & 0.34 & 0.29 & 0.27 & 0.35 & 0.32 & 0.31 & 0.40 & 0.40 & 0.40\\
& Qwen2.5-72B & 0.31 & 0.25 & 0.22 & 0.31 & 0.25 & 0.22 & 0.31 & 0.25 & 0.23\\
& DeepSeek-L70B & 0.39 & 0.34 & 0.36 & 0.35 & 0.32 & 0.31 & 0.36 & 0.32 & 0.31 \\
\cmidrule{2-11}
& Llama3.1-8B & 0.30 & 0.29 & 0.28 & 0.38 & 0.39 & 0.39 & 0.39 & 0.39 & 0.39 \\
& Qwen2.5-7B & 0.29 & 0.22 & 0.19 & 0.29 & 0.25 & 0.24 & 0.28 & 0.22 & 0.19\\
& DeepSeek-Q7B & 0.29 & 0.23 & 0.21 & 0.38 & 0.33 & 0.32 & 0.28 & 0.20 & 0.17 \\
& Mistral-7B & 0.30 & 0.23 & 0.21 & 0.34 & 0.30 & 0.28 & 0.36& 0.35 & 0.35 \\
\bottomrule
\end{tabular}
\label{main_colon}
\end{table*}
\begin{table*}[]
\centering
\caption{Main results for different LLM finetuning strategies on TRIALQA. cri represents the results for criterion eligibility prediction and int represents the results for interest reason prediction. }
\begin{tabular}{lc ccc ccc}
\toprule
\textbf{Dataset} &\textbf{Model} & \multicolumn{3}{c}{\textbf{Entry-level FT (cri/int)}} & \multicolumn{3}{c}{\textbf{User-level FT (cri/int)}} \\
\cmidrule(r){3-5} \cmidrule(r){6-8} 
 & & acc & MF1 & WF1 & acc & MF1 & WF1\\
\midrule
\multirow{4}{*}{Colon cancer}
& Roberta &  0.69/0.84 & 0.66/0.83 & 0.68/0.84 & 0.65/0.66 & 0.64/0.58 & 0.66/0.61  \\
& Llama3.1-8B & 0.62/0.61 & 0.56/0.37 & 0.60/0.46 & 0.68/0.66 & 0.68/0.59 & 0.68/0.62  \\
& Qwen2.5-7B & 0.57/0.58 & 0.46/0.37 & 0.53/0.43 & 0.62/0.50 & 0.61/0.50 & 0.63/0.50  \\
& DeepSeek-Q7B & 0.70/0.85 & 0.65/0.84 & 0.69/0.85 & 0.56/0.56 & 0.51/0.56 & 0.54/0.55 \\
& Mistral-7B & 0.64/0.84 & 0.62/0.83 & 0.64/0.84 & 0.55/0.55 & 0.53/0.55 & 0.55/0.54\\
\cmidrule{1-8}
\multirow{4}{*}{Prostate cancer} 
& Roberta & 0.68 &0.67 & 0.68 & 0.57 & 0.43 & 0.49\\
& Llama3.1-8B & 0.69 &0.69 & 0.69 & 0.57 & 0.43 & 0.49\\
& Qwen2.5-7B & 0.66 & 0.65 & 0.66 & 0.46 & 0.34 & 0.34\\
& DeepSeek-Q7B & 0.68 & 0.68 & 0.68 & 0.37 & 0.18 & 0.20\\
& Mistral-7B & 0.64 & 0.63 & 0.63 & 0.60 & 0.56 & 0.58\\
\bottomrule
\end{tabular}
\label{main_colon_ft}
\end{table*}

\subsection{RQ1: How do different LLM training and inference strategies perform on the smart trial recruitment task?}
In Table~\ref{main_colon} and Table~\ref{main_colon_ft}, we observe that, interestingly, for the direct prompting method on colon cancer, smaller language models tend to outperform larger ones. For example, Mistral-7B achieves the best performance among the seven LLMs; Llama3.1-8B and DeepSeek-Q7B outperform Llama3-70B and Qwen2.5-72B. This counterintuitive result may be attributed to the tendency of larger models to generate more verbose or unconstrained outputs, often including extraneous reasoning instead of strictly adhering to the instruction to provide only a final label. Such behavior can interfere with post-processing steps such as regex-based answer extraction. The consistent performance improvements observed with the ICL method--where few-shot examples help guide the model to follow the desired output format--further support this interpretation. Overall, compared to the colon cancer dataset, all models show worse performance in the prostate cancer dataset, suggesting that it is harder to identify CT participants for prostate cancer on Reddit.  Counterintuitively, self-consistency methods lead to performance degradation. This may result from the constrained output space (True, False, Unknown), which limits reasoning diversity. Consequently, sampling adds variability without yielding meaningful gains, ultimately lowering accuracy.

An additional noteworthy finding is the strong performance of Mistral-7B. On the colon cancer dataset, it surpasses all other models across the three inference strategies—Direct, ICL, and Self-Consistency—even outperforming larger models. This highlights Mistral-7B’s potential as an efficient and effective model for CT participant identification. One contributing factor may be its training on long contexts (up to 8K tokens), which aligns well with the long input sequences formed by aggregated user posts. On the prostate cancer dataset, Mistral-7B with Self-Consistency and ICL also achieves competitive performance. Interestingly, for both datasets, the non-LLM baseline NLI model RoBERTa (roberta-large-mnli) presents competitive performance compared to most LLMs in all three inference strategies. We hypothesize it is because these LLMs were not trained specifically on NLI tasks or explicitly designed for instruction following.
\begin{table*}[]
\centering
\caption{Accuracy across different criteria and interest reasons using the ICL method in the colon cancer dataset.}
\resizebox{0.8\linewidth}{!}{
\begin{tabular}{ccccccccccccccc}
\toprule
Models & c1 & c2 & c3 & c4 & c5 & c6 & c7 & c8 & r1 & r2 & r3 & r4 & r5 & r6 \\
\midrule
Llama3-70B & 0.34 & 0.51 & 0.40 & 0.52 & 0.36 & 0.98 & 0.50 & 0.45 & 0.61 & 0.65 & 0.42 & 0.78 & 0.74 & 0.57 \\
Qwen2.5-72B & 0.27 & 0.54 & 0.83 & 0.24 & 0.38 & 0.96 & 0.49 & 0.45 & 0.65 & 0.67 & 0.39 & 0.78 & 0.74 & 0.57\\
DeepSeek-L70B & 0.35 & 0.52 & 0.16 & 0.61 & 0.40 & 0.87 & 0.49 & 0.45 & 0.62 & 0.67 & 0.44 & 0.77 & 0.73 & 0.57 \\
\hline
Llama3.1-8B & 0.36 & 0.67 & 0.33 & 0.31 & 0.49 & 0.49 & 0.42 & 0.42 & 0.60 & 0.59 & 0.43 & 0.77 & 0.71 & 0.55 \\
Qwen2.5-7B & 0.33 & 0.47 & 0.80 & 0.24 & 0.35 & 0.98 & 0.46 & 0.44 & 0.65 & 0.67 & 0.34 & 0.77 & 0.73 & 0.56 \\
DeepSeek-Q7B & 0.28 & 0.36 & 0.85 & 0.26 & 0.37 & 0.97 & 0.48 & 0.46 & 0.65 & 0.67 & 0.37 & 0.78 & 0.74 & 0.57 \\
Mistral-7B & 0.32 & 0.60 & 0.32 & 0.69 & 0.40 & 0.99 & 0.50 & 0.48 & 0.63 & 0.68 & 0.56 & 0.76 & 0.70 & 0.57 \\
Average & 0.32 & 0.52 & 0.53 & 0.41 & 0.39 & 0.89 & 0.48 & 0.45 & 0.63 & 0.66 & 0.42 & 0.77 & 0.73 & 0.57 \\
\bottomrule
\end{tabular}
}
\label{icl}
\end{table*}
\begin{table}[!h]
\centering
\caption{Accuracy across different criteria and interest reasons using the ICL method in the prostate cancer dataset.}
\resizebox{1\linewidth}{!}{
\begin{tabular}{ccccccccc}
\toprule
Models & c1 & c2 & c3 & c4 & c5 & c6 & c7 & c8  \\
\midrule
Llama3-70B & 0.47 & 0.60 &  0.35  & 0.27 & 0.41 & 0.41 & 0.07 & 0.19 \\
Qwen2.5-72B & 0.27 & 0.54 & 0.83 & 0.24 & 0.38 & 0.96 & 0.49 & 0.45\\
DeepSeek-L70B &  0.49 & 0.61 &  0.47 & 0.25 & 0.35 & 0.39 & 0.08 & 0.19 \\
\hline
Llama3.1-8B &  0.46 &  0.50 &0.45 & 0.25 & 0.41 & 0.40 & 0.30 & 0.29\\
Qwen2.5-7B & 0.35 & 0.53  & 0.24 & 0.27 & 0.24 & 0.41 & 0.10 & 0.22 \\
DeepSeek-Q7B & 0.60 & 0.56 & 0.55 & 0.23 & 0.39 & 0.36 & 0.10 & 0.23 \\
Mistral-7B & 0.49 & 0.61 & 0.42 & 0.24 & 0.32 & 0.40 & 0.08& 0.18 \\
Average & 0.42 & 0.56 & 0.44 & 0.25 & 0.36 & 0.56 & 0.17 & 0.25 \\
\bottomrule
\end{tabular}
}
\label{proicl}
\end{table}

As shown in Table~\ref{main_colon_ft}, fine-tuning consistently improves performance over non–fine-tuned LLMs for both eligibility criteria and interest reason prediction tasks, indicating that user posts contain extractable signals about these attributes. Entry-level fine-tuning generally outperforms user-level fine-tuning, likely because the entry-level dataset is larger and more diverse. Similar patterns are observed with RoBERTa, which achieves either the best or near-best results. These findings suggest that, while LLMs exhibit certain capabilities, they are not necessarily superior to a smaller, more cost-effective NLI model for identifying potential CT participants on social media. In the following sections, we conduct in-depth analyses to explore possible reasons for these observations.

\subsection{RQ2: How do different LLM strategies perform on different eligibility criteria and interest reasons?}

To assess the varying difficulty of determining different eligibility criteria and interest reasons, we evaluate model performance using the ICL inference method as an example. LLMs tend to follow instructions more closely in ICL, enabling more meaningful analysis. 
\begin{table*}[!h]
\centering
\caption{Results on different classes for direct prompt and ICL on colon cancer data. Each entry represents precision, recall, and F1-score, respectively.}
\begin{tabular}{l|ccc|ccc}
\hline
\textbf{Model} & \multicolumn{3}{c|}{\textbf{Direct}} & \multicolumn{3}{c}{\textbf{ICL}} \\
& True & False & Unknown & True & False & Unknown \\
\hline
Llama3-70B     &   0.54/0.19/0.28  &  0.52/0.55/0.54   &  0.51/0.22/0.30    &  0.48/0.29/0.37   &  0.62/0.69/0.65   &   0.50/0.65/0.57   \\
Qwen2.5-72B  & 0.83/0.09/0.16  &  0.62/0.70/0.66   &  0.51/0.97/0.67   &   0.83/0.10/0.17  &  0.62/0.69/0.65   &   0.51/0.96/0.67 \\
DeepSeek-L70B   &  0.54/0.19/0.28   &   0.52/0.82/0.64  &  0.52/0.50/0.51   &   0.50/0.26/0.34  &   0.61/0.67/0.64  &  0.53/0.75/0.62   \\
\hline
Llama3.1-8B     &  0.55/0.29/0.38   &   0.53/0.69/0.60  &   0.50/0.57/0.53  &   0.52/0.29/0.37  &   0.50/0.74/0.60  &  0.52/0.45/0.48   \\
Qwen2.5-7B    &  0.70/0.10/0.18   &  0.61/0.69/0.65   &  0.51/0.92/0.65   &   0.65/0.06/0.11  &   0.61/0.70/0.65  &  0.51/0.94/0.66   \\
DeepSeek-Q7B    &  0.44/0.35/0.39   &  0.64/0.43/0.52   &  0.52/0.87/0.65   &  0.64/0.06/0.12   &  0.63/0.68/0.65   &   0.50/0.97/0.66  \\
Mistral-7B    &    0.67/0.14/0.23  &  0.64/0.66/0.65   &   0.52/0.97/0.67  &  0.57/0.36/0.44   &  0.64/0.66/0.65   &   0.53/0.74/0.62  \\
\hline
\end{tabular}
\label{tab:label}
\end{table*}
\begin{table*}[!h]
\centering
\caption{Results for different classes for direct prompt and ICL on prostate data. Each entry represents precision, recall, and F1-score, respectively.}
\begin{tabular}{l|ccc|ccc}
\hline
\textbf{Model} & \multicolumn{3}{c|}{\textbf{Direct}} & \multicolumn{3}{c}{\textbf{ICL}} \\
& True & False & Unknown & True & False & Unknown \\
\hline
Llama3-70B     &   0.56/0.25/0.35  &  0.37/0.05/0.09  &  0.29/0.92/0.44    &  0.66/0.28/0.39   &  0.41/0.11/0.17   &   0.27/0.83/0.41   \\
Qwen2.5-72B  &  0.81/0.15/0.25   &  0.35/0.04/0.07   &  0.27/0.96/0.42   &   0.87/0.17/0.28  &   0.38/0.03/0.05  &  0.26/0.96/0.41   \\
DeepSeek-L70B   &  0.44/0.69/0.54  &   0.32/0.23/0.27  &  0.29/0.17/0.22  &   0.70/0.29/0.41  &    0.40/0.09/0.14  &  0.28/0.87/0.42   \\
\hline
Llama3.1-8B     &  0.50/0.18/0.27   &   0.39/0.17/0.24  &   0.24/0.69/0.36  &   0.57/0.35/0.43  &   0.46/0.29/0.36  &  0.27/0.58/0.37   \\
Qwen2.5-7B  &0.75/0.11/0.19  &  0.41/0.03/0.06   &  0.25/0.96/0.40   &   0.79/0.13/0.23  &  0.36/0.10/0.15   &   0.25/0.85/0.38 \\  
DeepSeek-Q7B    &  0.35/0.28/0.31    &  0.56/0.00/0.00   &  0.26/0.75/0.39   &  0.57/0.49/0.53   &  0.54/0.02/0.05   &   0.28/0.75/0.40  \\
Mistral-7B    &    0.72/0.16/0.26  & 0.24/0.01/0.02   &   0.26/0.97/0.41  &  0.76/0.28/0.40   &  0.35/0.03/0.06   &   0.27/0.93/0.42  \\
\hline
\end{tabular}
\label{tab:labelp}
\end{table*}

\noindent\textbf{Results for colon cancer dataset}
Table~\ref{icl} presents the accuracy of various LLMs across individual criteria and reasons for interest in the colon cancer dataset. Overall, LLMs perform better on reasons-for-interest tasks, likely because they are simpler binary tasks, unlike the three-class eligibility criteria. Certain criteria—such as c2 (age between 18 and 95 years) and c6 (recent antibiotic use)—are more straightforward for LLMs to predict, likely because they involve explicit and easily identifiable textual cues. In contrast, other criteria requiring multi-hop reasoning pose greater challenges. Furthermore, model performance varies across criteria. For instance, DeepSeek-Q7B, Qwen2.5-7B, and Qwen2.5-72B excel on c3 (``Able to comprehend, sign, and date the written informed consent form''), whereas Mistral-7B, Qwen2.5-72B, and Llama3-70B perform better on c4 (``Able to provide informed consent in English''). These observations underscore the importance of aligning model capabilities with the specific reasoning requirements of each criterion.

Combining results from Figure~\ref{dis_colon} with those in Table \ref{icl}, we find that the overall performance of LLMs across criteria and reasons is moderate, with models achieving the highest scores on c6 and c3—criteria dominated by the ``Unknown'' label. This suggests that models tend to predict the ``Unknown'' label, as also shown in Section~\ref{class}. In contrast, criteria such as c1, c5, and r3–r6 exhibit lower model performance, while they have more balanced or skewed label distributions. The correlation between label skew and model performance highlights the limited generalizability of LLM-based methods: when data are imbalanced or contain fewer training signals for certain labels, prediction quality degrades. Notably, c3 and c6 have heavily imbalanced distributions, which likely contribute to inflated performance via bias toward the dominant class. These findings underscore the importance of accounting for label distributions when evaluating LLMs on real-world, imbalanced datasets.

\noindent\textbf{Results for prostate cancer dataset}
From Table~\ref{proicl} and Figure~\ref{dis_pro}, we observe that the prostate cancer dataset is more affected by the data distribution. Although c1 (``People with prostates $\geq$40 years of age'') suffers from label imbalance, with a high proportion of ``True'' labels, LLMs achieve relatively good performance on it because the criterion is direct and straightforward to predict. In contrast, most LLMs perform poorly on c7 and c8, which have low proportions of ``Unknown'' labels; LLMs tend to predict ``Unknown'' for harder questions. These findings indicate that LLM performance is influenced by both question difficulty and label distribution.

\subsection{RQ3: How do different LLMs perform on different classes?}
\label{class}
We further conduct in-depth analyses to evaluate the performance of LLMs on different classes, including ``True'' ``False'' and ``Unknown'' We present detailed results in Table~\ref{tab:label} and Table~\ref{tab:labelp}. We also include recall and precision to provide a more comprehensive analysis. This experiment focuses on Direct prompting and ICL, as self-consistency generally does not yield notable improvements. The results reveal that LLMs generally exhibit low recall for the ``True'' label, while achieving high recall for the ``Unknown'' label. This suggests that LLMs often default to predicting ``Unknown'' likely due to difficulty in extracting implicit information from user posts. In the prostate cancer dataset, we observe a similar finding that the recall for the ``False'' label is also very low.

When comparing ICL with direct prompting, we observe that larger models benefit more from ICL, showing greater improvements in precision, recall, and F1-score, particularly in the colon cancer dataset. In the prostate cancer dataset, most large and small models also benefit from it, especially for the prediction of the ``True'' label. This is likely because larger models are better equipped to handle longer and more complex inputs, allowing them to utilize the in-context examples effectively. In contrast, for the colon cancer dataset, smaller models show no gains from ICL and may even be hindered by the longer examples, resulting in degraded performance. While ICL can provide useful guidance for tackling more challenging problems, it offers little benefit—and may even introduce distraction—when applied to relatively easier questions in the colon cancer dataset.

\begin{table}[!h]
\centering
\caption{CoT performance on colon cancer and prostate cancer datasets. * means performance is improved by the CoT strategy compared to direct prompting.}
\resizebox{0.9\linewidth}{!}{
\begin{tabular}{lccc ccc}
\toprule
\textbf{Model} & \multicolumn{3}{c}{Colon cancer} & \multicolumn{3}{c}{Prostate cancer} \\
\cmidrule(lr){2-4} \cmidrule(lr){5-7}
 & Acc & MF1 & WF1 & Acc & MF1 & WF1 \\
\midrule
Llama3-70B & 0.54*& 0.50* & 0.51* & 0.22 & 0.24 & 0.26 \\
Qwen2.5-72B & 0.57 & 0.51* & 0.51* & 0.15 & 0.14 & 0.14 \\
DeepSeek-L70B & 0.48* & 0.46* & 0.47* & 0.37 & 0.35* & 0.40* \\
\midrule
Llama3.1-8B & 0.50 & 0.47 & 0.48 & 0.31* & 0.26 & 0.29 \\
Qwen2.5-7B & 0.56 & 0.48 & 0.48 & 0.14 & 0.14 & 0.13 \\
DeepSeek-Q7B & 0.41 & 0.38 & 0.38 & 0.42* & 0.33* & 0.42* \\
Mistral-7B & 0.52 & 0.47 & 0.48 & 0.35* & 0.29* & 0.35* \\
\bottomrule
\end{tabular}
}
\label{tab:cot}
\end{table}
\subsection{RQ4: Can the popular CoT strategy improve LLMs' performance?}
CoT has been shown to improve LLMs' performance across various tasks. To assess the performance gains introduced by the CoT strategy in our task, we apply CoT and present the results in Table~\ref{tab:cot}, reporting accuracy, macro-F1, and weighted-F1 for seven models on the two datasets. The results indicate that larger models tend to benefit more from the CoT strategy, particularly on the colon cancer dataset. The prostate cancer dataset appears more challenging, with lower overall scores and larger uncertainty, potentially due to more difficult questions and more complex question. These findings suggest that larger models possess stronger reasoning capabilities, whereas the reasoning ability of smaller models is constrained by their limited reasoning capacity. We observe that CoT yields smaller performance gains than ICL, likely because ICL is more straightforward while the required reasoning is harder to execute.
\subsection{RQ5: What are the common types of errors made by these LLMs?}
In this section, we examine common types of
errors made by these LLMs in the criteria eligibility prediction tasks. The analysis is based on the CoT results, as errors from other methods (Direct Prompt, ICL, Self-Consistency) are generally less informative. We identify three representative error categories (illustrated in \hyperref[box:error_examples]{\textit{Error Examples}}): incorrect conclusions from correct reasoning, incorrect reasoning, and confusion between ``No'' and ``Unknown'' labels. We find these error types across different models and criteria. These errors occur repeatedly in the generated answers, demonstrating that current LLMs still lack robust reasoning ability, the capacity to draw accurate conclusions from reasoning, and consistent instruction-following skills.
\section{Related work}
\subsection{CT recruitment}

Traditional recruitment methods, such as personal networking, medical system database and newspaper advertisements, have proven to be both costly and ineffective in reaching underrepresented populations \cite{heller2014strategies, friedman2013people, demeter2002selecting}. These methods often lack scalability, fail to engage potential participants meaningfully, and do not address the unique barriers faced by underrepresented communities. While current social media recruitment has expanded reach by leveraging the vast and diverse user base of social media platforms \cite{darmawan2020role, topolovec2016use}, it often relies on broadly targeted advertisements lacking a data-driven approach, leading to lower recruitment rates. Emerging AI-driven approaches have shown significant promise in recruiting CT participants \cite{guan2023cohortgpt, cai2021improving}. 
\begin{errorbox}\label{box:error_examples}
\textbf{Incorrect Result from Correct Reasoning} 

Criterion: The user was recently diagnosed with colorectal cancer (CRC) and is scheduled for a colectomy.

Label: Unknown

Reasoning: The writer's father was diagnosed with colon cancer, but there is no mention of the writer themselves being diagnosed or having a colectomy.

Answer based on reasoning: No
\vspace{0.1in}

\textbf{Incorrect Reasoning} 

Criterion: The user is between 18 and 95 years old.

Label: Unknown

Reasoning: The writer is between 18 and 95 years old.

Step-by-step explanation:
The sentence mentions that the writer is undergoing chemotherapy for colon cancer diagnosed in late August. Chemotherapy is typically a treatment for adults, and the context suggests the writer is an adult dealing with cancer treatment. Therefore, the writer is between 18 and 95 years old.

Answer based on reasoning: Yes

\vspace{0.1in}
\textbf{Confusion Between ``No'' and ``Unknown'' Labels}

Criterion: Documented germline pathogenic variant in known or suspected genes associated with prostate cancer risk.

Label: Unknown

Reasoning: No, the sentence does not provide information about the writer having a documented prostate cancer-related gene variant. So the answer is ``No.''

Answer based on reasoning: No
\end{errorbox}

However, current AI-driven approaches focus on optimizing prediction performance metrics like accuracy, without considering the fairness and interpretability of their designs \cite{chen2022visual, ezzati2020machine, miotto2015case, calaprice2020improving, beck2020artificial, gligorijevic2019optimizing, yaghy2022artificial, lu2024artificial}. Such approaches may inadvertently perpetuate biases and fail to adequately address the unique needs of underserved communities. Additionally, existing social media recruitment methods do not adequately address barriers like education and distrust among underrepresented groups, which are essential for progressing from awareness to enrollment in CT. While virtual agents have been utilized for CT screening and education \cite{ghosh2024real, krieger2022tailoring}, their potential for application in social media recruitment remains largely untapped. To address these limitations, we propose evaluating LLM performance in aligning user-generated social media posts with CT eligibility criteria and interest reason, using a dataset specifically designed to identify potential participants and annotated by our professional annotators.

\subsection{LLM-based question answering}
In the current LLM question answering (QA) framework, the model is typically provided with a question and an instruction to generate an answer. The LLM then produces a response based on its internal knowledge acquired during pretraining \cite{petroni2019language}. To enhance prediction accuracy, several methods have been developed. These include zero-shot prompting, where the question is directly input into the LLM; in-context learning or few-shot prompting, where a few example QA pairs are included in the prompt to guide the model \cite{brown2020language}; chain-of-thought (CoT) reasoning, which encourages the model to generate intermediate reasoning steps before arriving at the final answer \cite{wei2022chain}; self-consistency method determines the final answer by selecting the majority vote among the sampled answers. \cite{wang2022self}, and retrieval-augmented generation (RAG), which retrieves relevant external documents and uses them as additional context for answering the question \cite{lewis2020retrieval}. Besides these inference-time methods, LLMs can also be fine-tuned on QA datasets for the task to improve their performance \cite{wei2021finetuned}. A common way to do this is by adding a LoRA \cite{hu2022lora} adapter, which is a small extra set of trainable parameters, while keeping the rest of the original model frozen.

In our work, we curated a new dataset TRIALQA to evaluate LLMs' performance of direct prompting, ICL, CoT reasoning, self-consistency and fine-tuning on identifying potential CT participants on social media.

\section{Conclusion and Future Work}
This work investigates the use of LLMs for identifying potential CT participants on social media, a rich source of health-related information. To evaluate LLM capabilities for this task, we curate a novel dataset, TRAIQA, collected from Reddit. We conduct extensive experiments across diverse LLM architectures and advanced prompting strategies to assess their effectiveness. Our findings show substantial performance variation across models, with Mistral-7B performing comparatively well among the LLMs tested. However, overall performance remains suboptimal, indicating that current LLMs struggle with the complex, multi-hop reasoning often required in this domain. In many cases, the smaller, cost-efficient NLI model RoBERTa achieves the best results, highlighting the need for future research to improve LLM reasoning and generalization for real-world CT applications.

This study has several limitations. First, the dataset covers only two U.S.-based CTs, whereas recruitment requirements vary widely across trials and regions, leaving considerable room for expansion in future datasets. Second, our data source is limited to Reddit, which, despite its large pool of potential participants, may constrain both demographic diversity and the variety of language styles. Additionally, ethical constraints currently prevent us from releasing the dataset; we are working with Reddit to explore options for responsible data sharing. Future work should incorporate data from other platforms, such as Facebook, Patient.info, Doctissimo, and Onmeda, to enhance diversity and representativeness. Despite these limitations, our work provides the most comprehensive evaluation to date of LLM performance in online CT recruitment and introduces two datasets that can be used for downstream fine-tuning and inference. We believe this work can inform and advance the development of AI-driven methods for CT participant recruitment.

\bibliographystyle{plainnat}
\bibliography{abbrev}

\end{document}